\documentclass[11pt]{article}
\usepackage{times}
\usepackage{sidecap}
\usepackage{geometry}
\usepackage{floatrow}
\usepackage{float}
\usepackage{hyperref}
\usepackage{graphics,graphicx}
\usepackage{subcaption}
\usepackage{epsf,wrapfig}

\usepackage{natbib,bm}

\usepackage{paralist}
\let\olditem\item
\renewenvironment{thebibliography}[1]{%
  \section*{\refname}
  \let\par\relax
  \renewcommand{\item}[1][]{\olditem[\textbullet]}%
  \inparaenum}{\endinparaenum}

\newcounter{aff}

\DeclareGraphicsExtensions{.jpg,.pdf,.png,.eps}

\usepackage{caption}
\usepackage{xcolor}

\usepackage{tcolorbox}

\bibpunct{(}{)}{;}{a}{}{,}

\usepackage{soul}

\definecolor{darkcyan}{rgb}{0.0, 0.0, 0.7}

\definecolor{boxcolor}{RGB}{193, 229, 255}
\definecolor{figboxcolor}{RGB}{193, 229, 255}
\definecolor{boxcolorgray}{RGB}{227, 227, 227}
\definecolor{darkcyan}{rgb}{0.0, 0.0, 0.7}

\sethlcolor{boxcolorgray}

\newcommand{\my}{\mbox{$M_{\odot}$~yr$^{-1}$}}

\def\gtrsim{\lower 2pt \hbox{$\, \buildrel {\scriptstyle >}\over
{\scriptstyle \sim}\,$}}
\def\lesssim{\lower 2pt \hbox{$\, \buildrel {\scriptstyle <}\over
{\scriptstyle \sim}\,$}}

\def\approxlt{\lower.2em\hbox{$\buildrel < \over \sim$}}
\def\approxgt{\lower.2em\hbox{$\buildrel > \over \sim$}}

\def\msun{\rm M_{\odot}}

\def\be{\begin{equation}}
\def\ee{\end{equation}}
\def\bea{\begin{eqnarray*}}
\def\eea{\end{eqnarray*}}

\def\jref@jnl#1{{\rm#1}}
\def\aj{\jref@jnl{AJ}}			
\def\araa{\jref@jnl{ARA\&A}}		
\def\apj{\jref@jnl{ApJ}}			
\def\apjl{\jref@jnl{ApJ}}		
\def\apjs{\jref@jnl{ApJS}}		
\def\ao{\jref@jnl{Appl.~Opt.}}		
\def\apss{\jref@jnl{Ap\&SS}}		
\def\aap{\jref@jnl{A\&A}}		
\def\aapr{\jref@jnl{A\&A~Rev.}}		
\def\aaps{\jref@jnl{A\&AS}}		
\def\azh{\jref@jnl{AZh}}			
\def\baas{\jref@jnl{BAAS}}		
\def\jrasc{\jref@jnl{JRASC}}		
\def\memras{\jref@jnl{MmRAS}}		
\def\mnras{\jref@jnl{MNRAS}}	
\def\pasa{\jref@jnl{PASA}}
\def\pra{\jref@jnl{Phys.~Rev.~A}}	
\def\prb{\jref@jnl{Phys.~Rev.~B}}	
\def\prc{\jref@jnl{Phys.~Rev.~C}}	
\def\prd{\jref@jnl{Phys.~Rev.~D}}	
\def\pre{\jref@jnl{Phys.~Rev.~E}}	
\def\prl{\jref@jnl{Phys.~Rev.~Lett.}}	
\def\pasp{\jref@jnl{PASP}}		
\def\pasj{\jref@jnl{PASJ}}		
\def\qjras{\jref@jnl{QJRAS}}		
\def\skytel{\jref@jnl{S\&T}}		
\def\solphys{\jref@jnl{Sol.~Phys.}}	
\def\sovast{\jref@jnl{Soviet~Ast.}}	
\def\ssr{\jref@jnl{Space~Sci.~Rev.}}	
\def\zap{\jref@jnl{ZAp}}			
\def\nat{\jref@jnl{Nature}}		
\def\iaucirc{\jref@jnl{IAU~Circ.}}
\def\aplett{\jref@jnl{Astrophys.~Lett.}}
\def\apspr{\jref@jnl{Astrophys.~Space~Phys.~Res.}}
\def\bain{\jref@jnl{Bull.~Astron.~Inst.~Netherlands}}
\def\fcp{\jref@jnl{Fund.~Cosmic~Phys.}}
\def\gca{\jref@jnl{Geochim.~Cosmochim.~Acta}}
\def\grl{\jref@jnl{Geophys.~Res.~Lett.}}
\def\jcp{\jref@jnl{J.~Chem.~Phys.}}	
\def\jgr{\jref@jnl{J.~Geophys.~Res.}}	
\def\jqsrt{\jref@jnl{J.~Quant.~Spec.~Radiat.~Transf.}}
\def\memsai{\jref@jnl{Mem.~Soc.~Astron.~Italiana}}
\def\nphysa{\jref@jnl{Nucl.~Phys.~A}}
\def\physrep{\jref@jnl{Phys.~Rep.}}
\def\physscr{\jref@jnl{Phys.~Scr}}
\def\planss{\jref@jnl{Planet.~Space~Sci.}}
\def\procspie{\jref@jnl{Proc.~SPIE}}

\def\ion#1#2{#1$\;${\small\rm\@Roman{#2}}\relax}

\newcount\lecurrentfam
\def\LaTeX{\lecurrentfam=\the\fam \leavevmode L\raise.42ex
\hbox{$\fam\lecurrentfam\scriptstyle\kern-.3em A$}\kern-.15em\TeX}
\def\sizrpt{
(\fontname\the\font): em=\the\fontdimen6\font, ex=\the\fontdimen5\font
\typeout{
(\fontname\the\font): em=\the\fontdimen6\font, ex=\the\fontdimen5\font
}}
\definecolor{dark}{rgb}{ 0.34, 0.09, 0.49}

\def\rstar{$R_\star$}

\usepackage{fancyhdr,url}
\usepackage{transparent}
\usepackage{tablefootnote}
\usepackage{mathpazo}

\usepackage{hyperref}

\begin{document}

\noindent
\begin{raggedright}
{\Huge\bfseries Evolved Stars: Dust formation and mass loss \\[0.5em]
\Large The extended atmospheres and inner envelopes of evolved stars 
}\end{raggedright}\\

\noindent
\vspace{1em}

\noindent Luis Velilla-Prieto$^{1}$,
Wouter Vlemmings$^{2}$,
Guillermo Quintana-Lacaci$^{1}$\footnotemark
%
\footnotetext{E-mails: \url{l.velilla@csic.es}, \url{wouter.vlemmings@chalmers.se}, \url{guillermo.q@csic.es}}
\\
Javier Alcolea$^3$
Jaime Alonso-Hern\'andez$^4$,
Alain Baudry$^5$, 
Cristobal Bordiu$^6$, 
Ta{\"i}ssa Danilovich$^7$, 
Elvire De Beck$^2$, 
Leen Decin$^8$, 
Sandra Etoka$^9$,
J. Pablo Fonfría$^1$, 
Tomasz Kaminski$^{10}$,
Theo Khouri$^2$, 
Matthias Maercker$^2$,
Bannawit Pimpanuwat$^{11}$,
Raghvendra Sahai$^{12}$,
Carmen S{\'a}nchez Contreras$^4$, 
Miguel Santander-Garc\'ia$^3$,
Ramlal Unnikrishnan$^4$,
Marie Van de Sande$^{13}$


\vspace{1em}
{
\footnotesize{
\noindent 
$^1$ Institute of Fundamental Physics (IFF), CSIC, Serrano 123, 28006, Madrid, Spain \\
$^2$ Department of Physics and Astronomy, Chalmers University of Technology, SE-412 96 Gothenburg, Sweden \\
$^3$ Observatorio Astron\' omico Nacional, IGN/GNIG, MITMS, Spain \\
$^4$ Centro de Astrobiolog\'ia (CAB), CSIC-INTA, ESAC-campus, Camino Bajo del Castillo s/n, 28692, Villanueva de la Ca\~nada, Madrid, Spain \\
$^5$ Univ. of Bordeaux, L.A.B., France \\
$^6$ Instituto de Astrof\'isica de Andaluc\'ia, IAA-CSIC, Spain \\
$^7$ University of Newcastle, Australia \&\,Anton Pannekoek Institute for Astronomy, University of Amsterdam, The Netherlands \\
$^8$ KU Leuven, Belgium \\
$^9$ Jodrell Bank Centre for Astrophysics, The University of Manchester, Manchester M13 9PL, UK \\
$^{10}$ Nicolaus Copernicus Astronomical Center, ul. Rabia\'nska 8, 87-100 Toru\'n, Poland \\
$^{11}$National Astronomical Research Institute of Thailand (Public Organization), Chiang Mai 50180, Thailand\\
$^{12}$ Jet Propulsion Laboratory, California Institute of Technology, Pasadena, CA 91109, USA \\
$^{13}$ School of Physics, University College Dublin, Belfield, Dublin 4, Ireland \\
}
}
\begin{tcolorbox}[boxsep=3pt,left=0pt,right=0pt,top=1pt,bottom=1pt,colframe=boxcolor,colback=boxcolor]
\justify
\textbf{Abstract:} 
Evolved stars drive the chemical enrichment of the Universe, acting as primary producers of elemental products of nucleosynthesis and cosmic dust. 
Unraveling the exact role played by these objects requires empirical constraints on the interplay between pulsation, convection, magnetic fields, mass loss, and dust formation at small (milliarcsecond) scales, currently limited to only a handful of sources. 
Next-generation facilities, particularly the new ALMA2040, will enable systematic, spatially resolved studies of mass loss and dust formation across large samples of evolved stars, expanding from a few tens to hundreds of objects and providing the empirical constraints needed to understand the recycling and chemical enrichment of baryonic matter throughout the Universe.
\end{tcolorbox}

\newpage{}

\vspace{-0.7cm}
\section{Scientific context and open questions}
The final stages of stellar evolution play a crucial role in modern Astrophysics. 
Evolved stars are essential components of the cosmic matter cycle. 
During their late phases, stars expel material through stellar winds and explosive events, enriching the interstellar medium (ISM) with gas and dust. 
Through stellar nucleosynthesis — in the core, in shells during the asymptotic giant branch (AGB) phase, or in the explosive environments of supernovae (SN) — these stars produce the chemical elements and dust grains that constitute and become the seeds from which new stars and planets forms, with AGB stars producing over two thirds of the carbon and nitrogen in our Galaxy \citep[][and references therein]{2018A&ARv..26....1H}. 
Consequently, evolved stars drive galactic evolution and supply the building blocks necessary for the emergence of life.

A comprehensive description of stellar evolution, which underpins many of the most fundamental areas of astrophysical research, requires a thorough understanding of the physical processes that drive mass loss. 
Stellar evolution is governed by a range of internal mechanisms, including convection and magnetic processes, which, together with pulsations and other processes operating in the stellar atmospheres of evolved stars, play a fundamental role in driving mass loss and shaping stellar winds.
Mass loss is a highly complex phenomenon shaped by the interplay of convection, pulsations, dust formation, radiative and gravitational forces, and, often, companions and magnetic fields. 
Dust formation is closely connected to mass loss and stellar winds, as dust grains can form in the outflows where the physical conditions become favourable for condensation. 
The resulting dust can in turn contribute to the acceleration of the outflow, linking dust formation, mass loss, and the development of these stellar winds.
These relatively cold and dense stellar winds form circumstellar envelopes (CSE) that enshroud and hide the stars at optical and near infrared wavelengths.
	Nevertheless, their chemical complexity, gas kinematics, maser-type emission, magnetic fields, or dust properties (e.g. temperature, spectral index), as well as their temporal evolution, can be studied in the (sub-)millimeter wavelength range.
	This critical spectral window to investigate evolved stars has provided unprecedented advances in the understanding of the aforementioned physical processes \citep[e.g.,][]{2012Natur.490..232M,2020Sci...369.1497D,2023Natur.617..696V,2024Natur.633..323V}.

Significant progress is expected in our understanding of the late stages of stellar evolution, driven by advances in high-angular-resolution observations, multi-wavelength spectroscopy, and increasingly sophisticated models. 
However, many key questions cannot be addressed with current and next-decade instrumentation, particularly those requiring the direct characterization of the small-scale structures and temporal variability that shape stellar winds and ejecta originating from the atmosphere and inner envelope of evolved stars. 
Current-generation observations of continuum and molecular line emission at sub-millimeter wavelengths are limited to a small group of nearby stars, with spatial scales comparable to stellar sizes accessible for only about two dozen of them, while sensitivity and scheduling constraints can require prohibitively long observing times and make systematic temporal monitoring challenging.
Addressing these challenges will require a next generation of instrumentation, with which we finally aim to arrive at a full description of all physical processes and chemical pathways that underlie stellar mass loss and dust formation.

\subsection{Dust formation and circumstellar chemistry}
Evolved stars are the dominant sources of dust in galaxies \citep{2009MNRAS.397.1661V}. 
Dust composition depends on stellar chemistry: carbon-rich stars form grains mainly made of amorphous carbon and silicon carbide, while oxygen-rich stars form silicate- and Mg/Al-oxides-based grains \citep[][and references therein]{2001A&A...371..133F,2018A&ARv..26....1H}.
Observations in the sub-millimeter wavelength range enable to trace dust precursors and molecular species formed in the innermost regions of the CSE, such as AlO, TiO, SiO, SiC$_2$, and metal-bearing species, providing crucial insights into the early stages of dust formation, and tracing the regions where refractory elements condense, the so-called dust formation zone (DFZ).
This region extends from the stellar photosphere up to a few tens of stellar radii, and is characterized with high densities and low temperatures \citep[see e.g.,][]{2020A&A...637A..59A}. 
The advent of ALMA revealed for the first time a forest of lines arising from the DFZ in one of the best-studied evolved stars, IRC\,+\,10216 \citep{2013ApJ...778L..25C}.
	Most of the lines revealed in the partially observed DFZ spectrum of this object cannot be assigned to a known carrier, with tens to hundreds of unidentified lines associated with emission that appears at $\sim$\,1\,\rstar\ and disappears at $\sim$\,3\,\rstar, which is a strong indication that they should be related to dust precursors.
	Evolved stars can be used as spectroscopic laboratories as long as wide frequency ranges are observed to search for spectral patterns that identify molecules and new species.
	This technique has resulted in a very prolific source of molecular detections in space, particularly in evolved stars \citep[e.g.,][]{2019A&A...627L...4C,2025A&A...700L...6P}.
	However, current ALMA is not well suited for this method given its narrow spectral coverage per setup, requiring a prohibitive observing time to carry out a wide spectral scan of the DFZ even for the brightest and nearest evolved stars, resulting in fragmented results. 

Evolved stars also host a rich circumstellar chemistry, synthesizing complex molecules, including aromatic hydrocarbons, potential precursors of polycyclic aromatic hydrocarbons (PAHs), such as acetylene \citep{2008ApJ...673..445F}, and complex organic molecules (COMs), such as methanol, \citep{2017A&A...603L...2O} and their precursors \citep[e.g. HNCO,][]{2015A&A...575A..84V}.
This raises the possibility that evolved stars may not only supply the chemical ingredients later incorporated into the ISM, but also host environments where part of this chemical complexity develops locally \citep{2021MNRAS.501..491V}. 
It is possible that part of this material is incorporated into dust grains via adhesion in the innermost parts of the CSE \citep{1985ApJ...292..487J,1993AJ....105..595S}. 
Deep spectral surveys will therefore be essential for identifying the chemical pathways leading to the formation and destruction of relevant species, including COMs, their precursors, and molecular tracers of shocks and disk formation.
This is particularly important in regions where shocks and other dynamical phenomena may promote non-equilibrium chemistry through both gas-phase and gas-dust interactions, potentially involving icy mantles, as suggested by observations of sources such as OH\,231.8\,+\,4.2 \citep[][and references therein]{2018A&A...618A.164S}. 
Furthermore, elemental abundances and isotopic ratios derived from the molecules provide direct constraints on nucleosynthesis and stellar evolution \citep{2014PASA...31...30K}.

Despite the importance of the nucleation process, many aspects of dust formation and grain growth remain poorly understood, including condensation sequence, the role of shocks and convection, and coupling with mass loss. 
In particular, the transition from small molecular clusters to macroscopic grains remains observationally challenging due to the small spatial and rapid temporal scales involved, limiting detailed studies. 
Future high-angular-resolution and multi-epoch observations are thus essential to bridge the gap between theory and observations, and to give impetus to laboratory and theoretical work, essential to understand dust formation as a kinetic process \citep{2020A&A...637A..59A}.

\subsection{Convection and magnetic fields}
The transport of energy through convection plays a critical role during many stages of stellar evolution \citep[e.g.,][]{2017LRCA....3....1K}. 
Convection is best characterized in the Sun, where millions of convective cells on the surface—known as granules—can be directly observed \citep{2009LRSP....6....2N}. 
Evolved stars represent a promising laboratory for empirical studies of convection because of their extended convective envelopes and large apparent diameters. 
Since the size of convective cells is determined by the pressure scale height in the outer layers, and thus by the surface gravity of the star, only a small number of large granules are expected on the surfaces of giant and supergiant stars \citep{1975ApJ...195..137S}.
Observations with ALMA have provided evidence that such large convective cells indeed exist on nearby red supergiants (RSG), including Betelgeuse \citep[][]{2017A&A...602L..10O}. 
For AGB stars, interferometric observations with the VLTI have revealed signatures of convective activity \citep{2017A&A...600A.136P}, while the first direct measurements of convective cell size and timescale for the most nearby AGB star R Doradus have been obtained using ALMA \citep{2024Natur.633..323V}.
Despite their recognized importance for both mass-loss processes and stellar evolution, a comprehensive comparison between observed convective properties (such as cell sizes, lifetimes, velocities and the effect of magnetic fields) and those predicted by models, requires a next-generation instrument that provides higher-angular resolution and sensitivity than that of current ALMA. 
In current stellar evolution and population synthesis models, convection during the late stages of stellar evolution remains parametrized, with unverified assumptions as to their efficiency and impact on mass loss and those mentioned above, that directly influence predicted stellar lifetimes and nucleosynthetic yields. 
Consequently, uncertainties propagate into our interpretation of, e.g. JWST observations of distant galaxies \citep{2025NatAs...9..128L}. 
A robust understanding of convection in evolved stars is essential for advancing both stellar astrophysics and galactic evolution, requiring the ability to regularly monitor a volume-limited sample of Red Giant Branch (RGB) and AGB stars as well as RSGs with expected convective cell sizes of only a fraction of the stellar size, that would require spatial resolutions of $3-5$~milliarcseconds.  
Additionally, magnetic fields and their influence in the launching mechanisms of stellar winds are poorly characterized, although, recently, it has been shown that magnetic activity in evolved stars may play an important role as the underlying mechanism of random and episodic coronal-mass-ejection-like eruptions \citep{2015A&A...573L...1O,2023A&A...669A..56Q,2026arXiv260823695V}. However, since these results are based on individual objects, regular monitoring in full polarimetric mode, is also needed to confront models that include magnetic fields as a driver of stellar winds.

\subsection{Mass loss}
AGB mass-loss has traditionally been attributed to pulsation and radiation-driven dust winds \citep{2018A&ARv..26....1H}. 
However, high-resolution observations reveal non-continuous, fragmented outflows \citep[][and references therein]{2023Natur.617..696V,2025A&A...704A.341D,2026A&A...707A.162K}, challenging this paradigm and raising fundamental questions about the role played by companions, photospheric activity, magnetic fields, dust formation efficiencies and dynamics, as well as the coupling between dust grains and gas. 
Post-AGB mass-loss is especially critical for the transition to planetary nebulae (PN), characterized by fast, collimated mass ejections whose origin and powering mechanisms, including magnetic fields, are still under debate \citep[e.g.,][]{2001A&A...377..868B,2014MNRAS.440L..16B}.
In addition to molecular lines that trace neutral circumstellar regions, millimetre radio recombination lines (mmRRLs) offer a unique probe of the innermost regions ($<$150\,au) of pre-PNe \citep{2017A&A...603A..67S}.
To date, pre-PNe remain barely resolved given that they are very distant objetcs and the sizes of the regions where jets are launched require high-angular resolution of mas scale.
For evolved massive stars ($\gtrsim$\,8\,$\msun$), our knowledge of mass ejection is even more limited.
Despite their large stellar size, nearby massive stars are scarce, and only a few have been studied in molecular emission, which show very complex CSEs composed of, apparently, random ejections \citep[][and references therein]{2025A&A...698A.179D}.
Promising detections in yellow hypergiants, B[e] supergiants and luminous blue variables reveal dense, molecule-rich disks and asymmetric outflows that remain barely resolved in most cases, leaving their formation and shaping mechanisms (rotationally enhanced mass-loss, non-radial pulsations, binary interactions) an open question. 
Key goals, inaccessible with current instrumentation, relate, for example, to the influence of metallicity on the mass-loss processes \citep{2008A&A...486..497W}, making a detailed study of low-metallicity environments (e.g. the LMC and SMC) essential to refine models across different galactic contexts.
Currently, most studies base their mass-loss rate estimates on SED modelling, which introduces substantial uncertainties due to poorly constrained assumptions about the dust composition, grain size, and dust formation efficiency (i.e. the dust-to-gas ratio).
Only recent efforts have been put to detect CO emission in a scarce number of objetcs in the Magellanic Clouds with ALMA \citep[][]{2016A&A...596A..50G,2026arXiv260816456G}.
Additionally, interactions between stellar winds and the ISM, crucial for galactic chemical evolution as they have unknown impact on the grain size distribution of the dust, are still poorly characterized \citep[][and references therein]{2023AJ....165..229S}. 
Existing bow shock detections trace only a fraction of these interactions and high-sensitivity observations are needed to uncover faint, extended features and determine their properties.

\section{ALMA2040: Technical Requirements}
Addressing these open questions requires a next-generation mm/submillimeter interferometer with the following technical requirements. 
A $10\times$ increase in instantaneous spectral line sensitivity, compared to current ALMA $(\sim8\times$ compared to anticipated ALMA-Wide Sensitivity Upgrade, WSU), a $10\times$ increase in continuum sensitivity, compared to current ALMA ($\sim5\times$ compared to anticipated ALMA-WSU), 
and a $9\times$ smaller point-spread-function (PSF) area through $3\times$ longer baselines than ALMA-WSU,
and ultra-wide bandwidth, potentially reaching hundreds of GHz (e.g. $\geq$\,512\,GHz), which would represent a revolutionary advance in the field. 
These requirements can be achieved primarily by increasing the collecting area compared to ALMA-WSU and advancements in receiver technology, while at the same time providing exquisite image fidelity (through excellent $uv$-coverage) across a range of angular scales. 
Full polarization capabilities with polarization purity of $<0.1\%$ will also be highly desirable for detailed studies of the magnetic fields in the extended atmospheres at milliarcsecond scales. 
More flexible scheduling of different configurations, allowing access to high-angular resolutions on a more regular basis, should also be an important goal for ALMA, as current scheduling only allows for limited access to extended configurations for one month every other year.

Together with innovative developments in receiver technology such as simultaneous multi-band capabilities and ultra-wide bandwidths, these technical specifications will allow for 
\textit{i}) milliarcsecond angular resolution, which is essential to resolve convective cells and the innermost circumstellar regions where dust nucleates and winds are launched,
\textit{ii}) enhanced sensitivity, crucial for detecting faint continuum,  molecular line, and mmRRL emission from close and distant evolved stars, even in nearby galaxies, and
\textit{iii}) increased bandwidth and multi-band capabilities, which are key for spectral line surveys aiming at identifying multiple features of a given species and characterization of the spectral index thorugh continuum observations across the whole ALMA-accesible range,
\textit{iv}) time-domain monitoring, which is necessary to capture the variability of processes such as convection, operating on timescales of weeks to months.

\subsection{Expected impact}
A future ALMA2040 with capabilities as described above, would provide a major advance in the study of the extended atmopsheres, the DFZ, convection, and mass-loss processes in hundreds of stars at an unprecedented level: from current projects requiring about ten hours to conduct a search for a few species and their spatial distribution in the closest evolved stars, to comprehensive knowledge of a large number of stars in our Galaxy and in nearby galaxies, with access to wide spectral scans in a few hours. 
It is very plausible that ALMA2040 would be able to obtain a closer end-to-end characterization of dust precursors, convection, and stellar-wind properties.
In order to measure its expected impact, we present here two examples of the changes that the field would experience.

\begin{figure}[t!]
\centering
\vspace{0.1cm}
\includegraphics[width=0.7\textwidth]{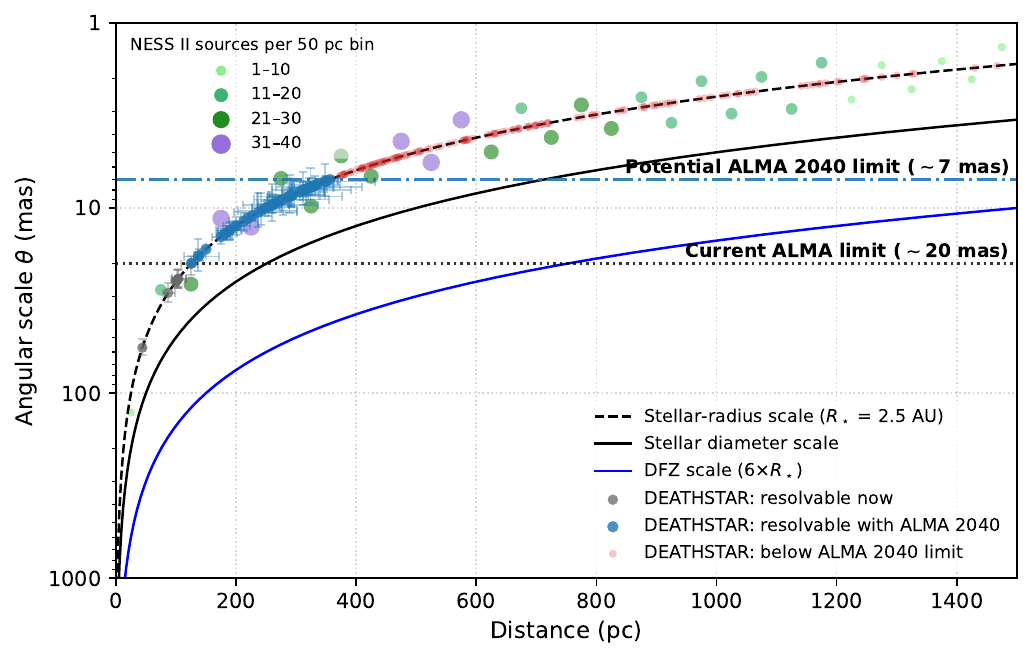}
\vspace{0.0cm}
\caption{\it Illustration of the improvement in evolved-star studies enabled by ALMA2040.
The population of AGB stars from \cite{2022A&A...667A..74A}, as coloured circles, are represented over the curve depicting the stellar-radius size, taken as 2.5\,au, as a function of distance to the source in pc. 
The volume-limited sample NESS \citep{2022MNRAS.512.1091S} is also represented as coloured circles of sources per distance bin according to the legend, with a small vertical offset from the curve for improved visualization.
The samples are restricted to sources within the ALMA observable declination range ($-70^\circ \leq \delta \leq +40^\circ$).
Horizontal lines represent the typical current ALMA high-angular resolution considering an average for all spectral bands ($\sim$\,20\,mas) and a future improvent by a factor of three.
Error bars in distance, according to \cite{2022A&A...667A..74A}, and their corresponding propagation to angular scales are shown for sources up to the 7\,mas\ limit.
For reference, the curves corresponding to a resolution equivalent to the AGB stellar diameter and to five stellar radii, required for proper spatial sampling of the DFZ, are shown.
\vspace{-0.4cm}}
\label{fig1}
\end{figure}

In Figure\,\ref{fig1}, we show the technical feasibility, in terms of achievable spatial resolution, of observational projects aiming to resolve evolved stars at different spatial scales and the expected leap forward an upgraded ALMA will offer.
We estimated the improvement based on the distances of the stars that can be observed from ALMA in the volume-limited sample Nearby Evolved Star Survey \citep[NESS,][]{2022MNRAS.512.1091S}  and in the ALMA DEATHSTAR project sample \citep{2022A&A...667A..74A}, which is not complete but includes a representative reference catalog of nearby AGB stars. 
Average value for the current spatial resolution and an improvement by a factor of $\sim$\,3, considering stellar-size scale resolution, with no additional limitations on sensitivities, have been taken into account for this approximation. 
We obtain an improvement of a factor $\sim$\,5 in the number of resolvable sources from both the NESS and DEATHSTAR samples (from 32 sources to 149 sources with ALMA2040, in the case of NESS, and from 6 to 31 in the case of DEATHSTAR).
A similar situation applies to the much scarcer population of evolved massive stars, for which only a few nearby objects can currently be studied at very high angular resolution despite their large stellar radii \citep{2018A&A...609A..67K}.
Furthermore, using the Mid-infrared stellar Diameters and Fluxes compilation Catalogue \citep[MDFC,][]{2019MNRAS.490.3158C} and considering giants and supergiants with infrared excesses or extended emission, an improvement by a factor of $\sim$\,10 could be achieved.
This value likely represents an upper limit, since the selected sample may include objects that are not evolved stars.

For mass-loss studies and investigations of the DFZ, the combination of improved angular resolution and sensitivity will make spatially resolved studies  feasible for substantially larger samples of evolved stars, allowing a more systematic characterization of its spatial structure and physical conditions. 
This will range from abundant molecules directly involved in dust-formation chemistry, for example Si--C-bearing species, such as SiC$_2$, Si$_2$C, or Si$_3$C, other silicates-related species such as SiO, to metal-bearing molecules relevant to the formation of observationally constrained condensates, such as MgS and TiC, which have not been detected yet \citep{2020A&A...637A..59A}.
In fact, we do not know which molecules act as seeds of the dust grains and more sensitive observations with larger bandwidth would be critical to e.g. search for spectral patterns for the characterization of all the unassigned detected features, which are probably important for the dust formation processes, that we currently have in the few but best studied objects, such as IRC+10216 \citep{2013ApJ...778L..25C}.

\begin{figure}[t!]
\centering
\vspace{0.1cm}
\includegraphics[width=0.85\textwidth]{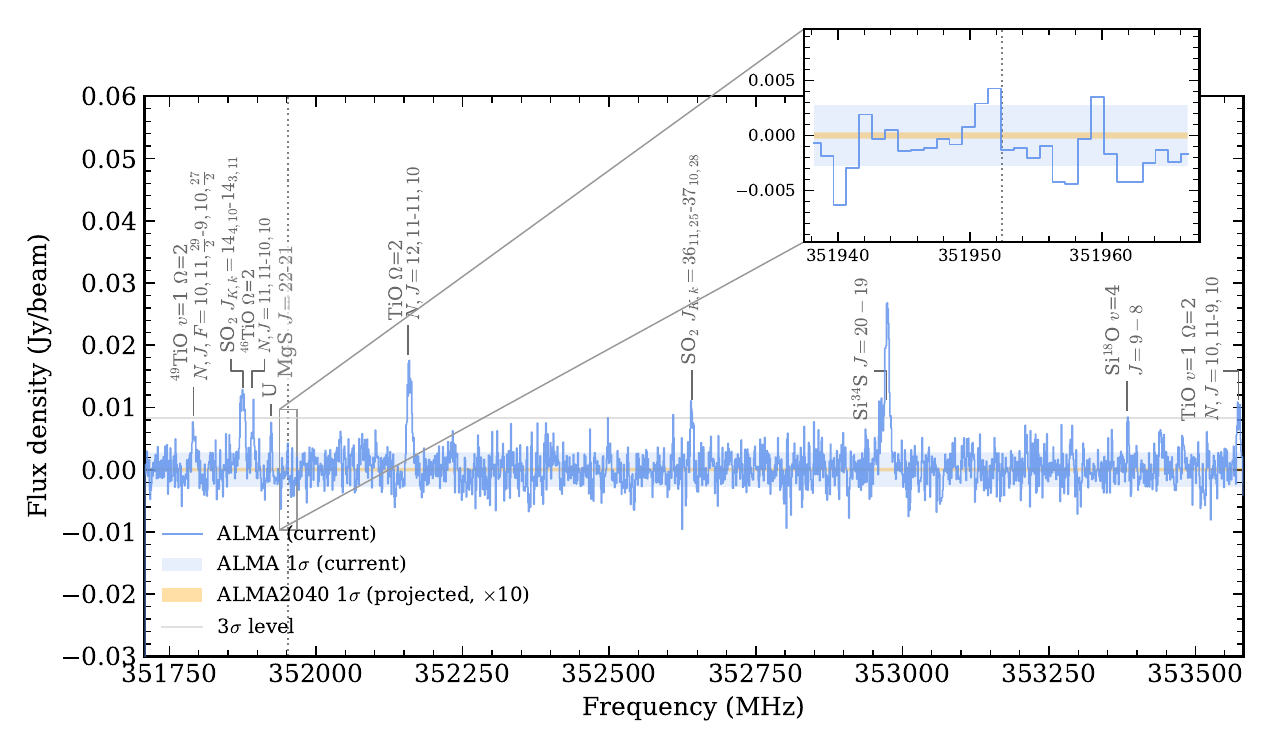}
\vspace{0.0cm} 
\caption{\it ALMA spectrum of the O-rich AGB star IK\,Tau extracted from a region of $\sim$\,25\,mas, which approximately corresponds to the beam size, centred on the stellar position and with a $\sim$\,1\,km\,s$^{-1}$ spectral resolution. Current and future sensitivity limits are shown for reference. The inset shows the spectrum around the frequency where the MgS $J$\,=\,22\,--\,21 line lies, which is not detected. In order to confirm if the peak close to the frequency of the MgS line, which reaches $\sim$\,4\,mJy\,beam$^{-1}$, is a spectral feature with $\mathrm{S/N}>3$, $\sim$\,6\,h of observing time would be required. ALMA2040 would reach such sensitivity goal in less than 5\,min. The current observations in the box were obtained in $\sim$\,1.5\,h of observing time in C-7 configuration. 
\vspace{-0.4cm}}
\label{fig2}
\end{figure}

In Figure\,\ref{fig2}, we show the ALMA spectrum of the well-known M-type star IK\,Tau, a close-by source with high mass-loss rate \citep[261\,pc\ and $\sim$\,5\,$\times$\,10$^{-6}$\,\my,][]{2026A&A...711A.159M}, as observed with high-angular resolution. 
Although the target was observed with a different scientific scope for $\sim$\,1.5\,h with configuration C-7, it covered the frequency where the MgS $J$\,=\,22--21 line lies. 
This molecule is predicted to be the most abundant Mg-bearing molecule around oxygen rich AGB stars, and potential gas-phase precursor of Mg-rich silicate dust \citep{2020A&A...637A..59A}. 
A complete and very sensitive ($\sim$\,1\,mJy\,beam$^{-1}$) detection experiment, would require about $\sim$\,15--20\,h\ in this target since multiple lines would have to be detected in order to confirm its presence. 
Spatially resolving the emission may further increase the observing time in order to properly characterize the radial distance from the central star, at which MgS appears or disappears from the gas-phase.
A similar observational effort would be needed, for example, to test if atomic carbon could be detected in the extended atmosphere of a single close-by carbon star, through its 492\,GHz line, or Al-bearing species, which have been detected in only $\sim$\,16\,stars, preventing us from a robust characterization of these important chemical species related to dust grains.
This is only a sample of what can be done for one particular species, such as MgS, but similar efforts would be required for the characterization of other metal-bearing molecules, such as Al-, Ti-, or Si-bearing molecules.
Thus, molecular transitions that currently require prohibitively long integrations could become accessible within standard observing programmes, transforming detailed studies of a few nearby prototypes into systematic surveys of evolved-star populations.
At the highest frequencies, ALMA2040 could even approach angular resolutions corresponding to spatial scales of a few hundred au at the distance of the Magellanic Clouds, opening the possibility of spatially resolved studies of individual circumstellar envelopes beyond the Milky Way and enabling systematic studies of metallicity-dependent effects on mass loss, dust formation, and circumstellar chemistry.

\section{Synergies with current and future facilities} 
Several current and future facilities will provide important synergies with ALMA across wavelengths and physical scales. 
Infrared facilities are also suited to investigate the chemical composition and properties of the circumstellar gas and dust, as well as the icy mantles. 
These are particularly relevant to study the evolved stellar population in low-metallicity galaxies, such as the Magellanic Clouds. 
The James Webb Space Telescope (JWST) has revealed results of paramount importance for the understanding of the growth of chemical complexity and circumstellar morphology in evolved stars \citep[e.g.,][]{2022NatAs...6.1421D,Li_2026}.
The PRobe far-Infrared Mission for Astrophysics (PRIMA), which has recently advanced to the preliminary design and technological development stage, would have strong synergies with future ALMA to study the life cycle of matter and the fundamental role played by evolved stars in the formation of dust and increasing chemical complexity.

The optical and near-infrared regime, in particular the Extremely Large Telescope (ELT),  will be essential to investigate variability and stellar photospheres and connect these studies with the formation of structures in the extended atmosphere and circumstellar envelopes of evolved stars.

At radio wavelengths, SKA and the future ngVLA will offer complementary capabilities to study maser emission, ionised gas, radio continuum, and magnetic fields. 
In particular, ALMA observations of (sub-)mm maser lines, in combination with cm-wave maser observations with SKA and the future ngVLA, will provide access to different maser-emitting regions across the circumstellar envelope, from the innermost regions close to the star to the outer envelope.
SKA will also be essential to study H\,I emission, providing a fundamental synergy with CO and mass-loss studies with ALMA.
SKA and ngVLA continuum observations will also provide an independent probe of mass-loss rates, which combined with estimates from (sub-)mm emission, can offer insights into wind structure and clumping.
Together, these facilities will help connect the molecular and dusty envelope with the innermost regions, providing important constraints on the physical mechanisms driving mass loss, including convection, shocks and dust formation, as well as help us to enlarge the list of the molecules detected in space, in particular, for heavy molecules (e.g. PAHs). 

Wide-field survey facilities such as Rubin, Euclid and SPHEREx will provide complementary information on variability and on large populations of evolved stars, extending studies beyond the relatively small samples that can be targeted at high angular resolution. 
These surveys will be particularly valuable for placing detailed ALMA studies of individual objects into a broader evolutionary and environmental context, including the effects of metallicity and stellar evolution.

Only anchored by an instrument like ALMA2040, that provides the necessary sensitive and high-angular observations of the extended atmospheres and dust formation region, will these facilities realise their full potential by enabling a multi-wavelength and multi-scale view of mass loss, from the stellar photosphere and the formation of dust to the molecular envelope, maser-emitting regions and the interaction with the surrounding medium.

\section*{Acknowledgements}
This work was developed as part of the broader ALMA2040 community effort. We acknowledge the Lorentz Center and NOVA for supporting the community workshops and coordination activities that helped initiate and shape the ALMA2040 White Paper process. We are grateful to Dr. Pierre Cox and Dr. Jos\'e Cernicharo for their careful reading of the paper and for their detailed suggestions that helped us to improve the manuscript. 
L.V.P. acknowledges grants RYC2023-045648-I and PID2025-169994NB-I00 funded by MICIU/AEI/10.13039/501100011033, by "ERDF/EU", and FSE+. W.V., and T.K. acknowledge support from the Olle Engkvist Foundation through grant 229-0368.
R.S.’s contribution to the research described here was carried out at the Jet Propulsion Laboratory, California Institute of Technology, under a contract with NASA (80NM0018D0004).
AI tools were used for grammatical and spelling checks and figure formatting.\\

\bibliographystyle{apj_mod}
{\bibliography{bibliography}}

@ARTICLE{2014PASA...31...30K,
       author = {{Karakas}, Amanda I. and {Lattanzio}, John C.},
        title = "{The Dawes Review 2: Nucleosynthesis and Stellar Yields of Low- and Intermediate-Mass Single Stars}",
      journal = {\pasa},
         year = 2014,
        month = jul,
       volume = {31},
          eid = {e030},
        pages = {e030},
          doi = {10.1017/pasa.2014.21},
archivePrefix = {arXiv},
       eprint = {1405.0062},
 primaryClass = {astro-ph.SR},
       adsurl = {https://ui.adsabs.harvard.edu/abs/2014PASA...31...30K}
}

@ARTICLE{2009LRSP....6....2N,
       author = {{Nordlund}, {\r{A}}ke and {Stein}, Robert F. and {Asplund}, Martin},
        title = "{Solar Surface Convection}",
      journal = {Living Reviews in Solar Physics},
         year = 2009,
        month = dec,
       volume = {6},
       number = {1},
          eid = {2},
        pages = {2},
          doi = {10.12942/lrsp-2009-2},
       adsurl = {https://ui.adsabs.harvard.edu/abs/2009LRSP....6....2N}
}

@ARTICLE{2024Natur.633..323V,
       author = {{Vlemmings}, Wouter and {Khouri}, Theo and {Bojnordi Arbab}, Behzad and {De Beck}, Elvire and {Maercker}, Matthias},
        title = "{One month convection timescale on the surface of a giant evolved star}",
      journal = {\nat},
         year = 2024,
        month = sep,
       volume = {633},
       number = {8029},
        pages = {323-326},
          doi = {10.1038/s41586-024-07836-9},
archivePrefix = {arXiv},
       eprint = {2409.06785},
 primaryClass = {astro-ph.SR},
       adsurl = {https://ui.adsabs.harvard.edu/abs/2024Natur.633..323V}
}

@ARTICLE{2025NatAs...9..128L,
       author = {{Lu}, Shiying and {Daddi}, Emanuele and {Maraston}, Claudia and {Dickinson}, Mark and {Haro}, Pablo Arrabal and {Gobat}, Raphael and {Renzini}, Alvio and {Giavalisco}, Mauro and {Bagley}, Micaela B. and {Calabr{\`o}}, Antonello and {Cheng}, Yingjie and {de la Vega}, Alexander and {D'Eugenio}, Chiara and {Elbaz}, David and {Finkelstein}, Steven L. and {G{\'o}mez-Guijarro}, Carlos and {Gu}, Qiusheng and {Hathi}, Nimish P. and {Huertas-Company}, Marc and {Kartaltepe}, Jeyhan S. and {Koekemoer}, Anton M. and {Henry}, Aur{\'e}lien and {Lyu}, Yipeng and {Magnelli}, Benjamin and {Mobasher}, Bahram and {Papovich}, Casey and {Pirzkal}, Nor and {Rich}, R. Michael and {Tacchella}, Sandro and {Yung}, L.~Y. Aaron},
        title = "{Strong spectral features from asymptotic giant branch stars in distant quiescent galaxies}",
      journal = {Nature Astronomy},
         year = 2025,
        month = jan,
       volume = {9},
        pages = {128-140},
          doi = {10.1038/s41550-024-02391-9},
archivePrefix = {arXiv},
       eprint = {2403.07414},
 primaryClass = {astro-ph.GA},
       adsurl = {https://ui.adsabs.harvard.edu/abs/2025NatAs...9..128L}
}

@ARTICLE{2001A&A...377..868B,
       author = {{Bujarrabal}, V. and {Castro-Carrizo}, A. and {Alcolea}, J. and {S{\'a}nchez Contreras}, C.},
        title = "{Mass, linear momentum and kinetic energy of bipolar flows in protoplanetary nebulae}",
      journal = {\aap},
         year = 2001,
        month = oct,
       volume = {377},
        pages = {868-897},
          doi = {10.1051/0004-6361:20011090},
       adsurl = {https://ui.adsabs.harvard.edu/abs/2001A&A...377..868B}
}

@ARTICLE{1975ApJ...195..137S,
       author = {{Schwarzschild}, M.},
        title = "{On the scale of photospheric convection in red giants and supergiants.}",
      journal = {\apj},
         year = 1975,
        month = jan,
       volume = {195},
        pages = {137-144},
          doi = {10.1086/153313},
       adsurl = {https://ui.adsabs.harvard.edu/abs/1975ApJ...195..137S}
}

@ARTICLE{2020A&A...637A..59A,
       author = {{Ag{\'u}ndez}, M. and {Mart{\'\i}nez}, J.~I. and {de Andres}, P.~L. and {Cernicharo}, J. and {Mart{\'\i}n-Gago}, J.~A.},
        title = "{Chemical equilibrium in AGB atmospheres: successes, failures, and prospects for small molecules, clusters, and condensates}",
      journal = {\aap},
         year = 2020,
        month = may,
       volume = {637},
          eid = {A59},
        pages = {A59},
          doi = {10.1051/0004-6361/202037496},
archivePrefix = {arXiv},
       eprint = {2004.00519},
 primaryClass = {astro-ph.SR},
       adsurl = {https://ui.adsabs.harvard.edu/abs/2020A&A...637A..59A}
}

@ARTICLE{2009MNRAS.397.1661V,
       author = {{Valiante}, Rosa and {Schneider}, Raffaella and {Bianchi}, Simone and {Andersen}, Anja C.},
        title = "{Stellar sources of dust in the high-redshift Universe}",
      journal = {\mnras},
         year = 2009,
        month = aug,
       volume = {397},
       number = {3},
        pages = {1661-1671},
          doi = {10.1111/j.1365-2966.2009.15076.x},
archivePrefix = {arXiv},
       eprint = {0905.1691},
 primaryClass = {astro-ph.CO},
       adsurl = {https://ui.adsabs.harvard.edu/abs/2009MNRAS.397.1661V}
}

@ARTICLE{2023Natur.617..696V,
       author = {{Velilla-Prieto}, L. and {Fonfr{\'\i}a}, J.~P. and {Ag{\'u}ndez}, M. and {Castro-Carrizo}, A. and {Gu{\'e}lin}, M. and {Quintana-Lacaci}, G. and {Cherchneff}, I. and {Joblin}, C. and {McCarthy}, M.~C. and {Mart{\'\i}n-Gago}, J.~A. and {Cernicharo}, J.},
        title = "{Atmospheric molecular blobs shape up circumstellar envelopes of AGB stars}",
      journal = {\nat},
         year = 2023,
        month = may,
       volume = {617},
       number = {7962},
        pages = {696-700},
          doi = {10.1038/s41586-023-05917-9},
archivePrefix = {arXiv},
       eprint = {2412.04027},
 primaryClass = {astro-ph.SR},
       adsurl = {https://ui.adsabs.harvard.edu/abs/2023Natur.617..696V}
}

@ARTICLE{2017LRCA....3....1K,
       author = {{Kupka}, Friedrich and {Muthsam}, Herbert J.},
        title = "{Modelling of stellar convection}",
      journal = {Living Reviews in Computational Astrophysics},
         year = 2017,
        month = jul,
       volume = {3},
       number = {1},
          eid = {1},
        pages = {1},
          doi = {10.1007/s41115-017-0001-9},
       adsurl = {https://ui.adsabs.harvard.edu/abs/2017LRCA....3....1K}
}

@ARTICLE{2008A&A...486..497W,
       author = {{Wachter}, A. and {Winters}, J.~M. and {Schr{\"o}der}, K.-P. and {Sedlmayr}, E.},
        title = "{Dust-driven winds and mass loss of C-rich AGB stars with subsolar metallicities}",
      journal = {\aap},
         year = 2008,
        month = aug,
       volume = {486},
       number = {2},
        pages = {497-504},
          doi = {10.1051/0004-6361:200809893},
archivePrefix = {arXiv},
       eprint = {0805.3656},
 primaryClass = {astro-ph},
       adsurl = {https://ui.adsabs.harvard.edu/abs/2008A&A...486..497W}
}

@ARTICLE{2001A&A...371..133F,
       author = {{Ferrarotti}, A.~S. and {Gail}, H.-P.},
        title = "{Mineral formation in stellar winds. II. Effects of Mg/Si abundance variations on dust composition in AGB stars}",
      journal = {\aap},
         year = 2001,
        month = may,
       volume = {371},
        pages = {133-151},
          doi = {10.1051/0004-6361:20010316},
       adsurl = {https://ui.adsabs.harvard.edu/abs/2001A&A...371..133F}
}

@ARTICLE{2017A&A...602L..10O,
       author = {{O'Gorman}, E. and {Kervella}, P. and {Harper}, G.~M. and {Richards}, A.~M.~S. and {Decin}, L. and {Montarg{\`e}s}, M. and {McDonald}, I.},
        title = "{The inhomogeneous submillimeter atmosphere of Betelgeuse}",
      journal = {\aap},
         year = 2017,
        month = jun,
       volume = {602},
          eid = {L10},
        pages = {L10},
          doi = {10.1051/0004-6361/201731171},
archivePrefix = {arXiv},
       eprint = {1706.06021},
 primaryClass = {astro-ph.SR},
       adsurl = {https://ui.adsabs.harvard.edu/abs/2017A&A...602L..10O}
}

@ARTICLE{2014MNRAS.440L..16B,
       author = {{Blackman}, E.~G. and {Lucchini}, S.},
        title = "{Using kinematic properties of pre-planetary nebulae to constrain engine paradigms.}",
      journal = {\mnras},
         year = 2014,
        month = may,
       volume = {440},
        pages = {L16-L20},
          doi = {10.1093/mnrasl/slu001},
archivePrefix = {arXiv},
       eprint = {1312.5372},
 primaryClass = {astro-ph.SR},
       adsurl = {https://ui.adsabs.harvard.edu/abs/2014MNRAS.440L..16B}
}

@ARTICLE{2015A&A...575A..84V,
       author = {{Velilla Prieto}, L. and {S{\'a}nchez Contreras}, C. and {Cernicharo}, J. and {Ag{\'u}ndez}, M. and {Quintana-Lacaci}, G. and {Alcolea}, J. and {Bujarrabal}, V. and {Herpin}, F. and {Menten}, K.~M. and {Wyrowski}, F.},
        title = "{New N-bearing species towards OH 231.8+4.2. HNCO, HNCS, HC$_{3}$N, and NO}",
      journal = {\aap},
         year = 2015,
        month = mar,
       volume = {575},
          eid = {A84},
        pages = {A84},
          doi = {10.1051/0004-6361/201424768},
archivePrefix = {arXiv},
       eprint = {1412.2074},
 primaryClass = {astro-ph.GA},
       adsurl = {https://ui.adsabs.harvard.edu/abs/2015A&A...575A..84V}
}

@ARTICLE{2017A&A...600A.136P,
       author = {{Paladini}, C. and {Klotz}, D. and {Sacuto}, S. and {Lagadec}, E. and {Wittkowski}, M. and {Richichi}, A. and {Hron}, J. and {Jorissen}, A. and {Groenewegen}, M.~A.~T. and {Kerschbaum}, F. and {Verhoelst}, T. and {Rau}, G. and {Olofsson}, H. and {Zhao-Geisler}, R. and {Matter}, A.},
        title = "{The VLTI/MIDI view on the inner mass loss of evolved stars from the Herschel MESS sample}",
      journal = {\aap},
         year = 2017,
        month = apr,
       volume = {600},
          eid = {A136},
        pages = {A136},
          doi = {10.1051/0004-6361/201527210},
archivePrefix = {arXiv},
       eprint = {1701.05407},
 primaryClass = {astro-ph.SR},
       adsurl = {https://ui.adsabs.harvard.edu/abs/2017A&A...600A.136P}
}

@ARTICLE{2017A&A...603A..67S,
       author = {{S{\'a}nchez Contreras}, C. and {B{\'a}ez-Rubio}, A. and {Alcolea}, J. and {Bujarrabal}, V. and {Mart{\'\i}n-Pintado}, J.},
        title = "{A pilot search for mm-wavelength recombination lines from emerging ionized winds in pre-planetary nebulae candidates}",
      journal = {\aap},
         year = 2017,
        month = jul,
       volume = {603},
          eid = {A67},
        pages = {A67},
          doi = {10.1051/0004-6361/201730385},
archivePrefix = {arXiv},
       eprint = {1704.01773},
 primaryClass = {astro-ph.SR},
       adsurl = {https://ui.adsabs.harvard.edu/abs/2017A&A...603A..67S}
}

@ARTICLE{2017A&A...603L...2O,
       author = {{Olofsson}, H. and {Vlemmings}, W.~H.~T. and {Bergman}, P. and {Humphreys}, E.~M.~L. and {Lindqvist}, M. and {Maercker}, M. and {Nyman}, L. and {Ramstedt}, S. and {Tafoya}, D.},
        title = "{First detection of methanol towards a post-AGB object, HD 101584}",
      journal = {\aap},
         year = 2017,
        month = jul,
       volume = {603},
          eid = {L2},
        pages = {L2},
          doi = {10.1051/0004-6361/201731170},
archivePrefix = {arXiv},
       eprint = {1706.08254},
 primaryClass = {astro-ph.SR},
       adsurl = {https://ui.adsabs.harvard.edu/abs/2017A&A...603L...2O}
}

@ARTICLE{2022A&A...667A..74A,
       author = {{Andriantsaralaza}, M. and {Ramstedt}, S. and {Vlemmings}, W.~H.~T. and {De Beck}, E.},
        title = "{Distance estimates for AGB stars from parallax measurements}",
      journal = {\aap},
         year = 2022,
        month = nov,
       volume = {667},
          eid = {A74},
        pages = {A74},
          doi = {10.1051/0004-6361/202243670},
archivePrefix = {arXiv},
       eprint = {2209.03906},
 primaryClass = {astro-ph.SR},
       adsurl = {https://ui.adsabs.harvard.edu/abs/2022A&A...667A..74A}
}

@ARTICLE{2018A&A...609A..67K,
       author = {{Kervella}, Pierre and {Decin}, Leen and {Richards}, Anita M.~S. and {Harper}, Graham M. and {McDonald}, Iain and {O'Gorman}, Eamon and {Montarg{\`e}s}, Miguel and {Homan}, Ward and {Ohnaka}, Keiichi},
        title = "{The close circumstellar environment of Betelgeuse. V. Rotation velocity and molecular envelope properties from ALMA}",
      journal = {\aap},
         year = 2018,
        month = jan,
       volume = {609},
          eid = {A67},
        pages = {A67},
          doi = {10.1051/0004-6361/201731761},
archivePrefix = {arXiv},
       eprint = {1711.07983},
 primaryClass = {astro-ph.SR},
       adsurl = {https://ui.adsabs.harvard.edu/abs/2018A&A...609A..67K}
}

@ARTICLE{2018A&A...618A.164S,
       author = {{S{\'a}nchez Contreras}, C. and {Alcolea}, J. and {Bujarrabal}, V. and {Castro-Carrizo}, A. and {Velilla Prieto}, L. and {Santander-Garc{\'\i}a}, M. and {Quintana-Lacaci}, G. and {Cernicharo}, J.},
        title = "{Through the magnifying glass: ALMA acute viewing of the intricate nebular architecture of OH 231.8+4.2}",
      journal = {\aap},
         year = 2018,
        month = oct,
       volume = {618},
          eid = {A164},
        pages = {A164},
          doi = {10.1051/0004-6361/201833632},
archivePrefix = {arXiv},
       eprint = {1809.00619},
 primaryClass = {astro-ph.SR},
       adsurl = {https://ui.adsabs.harvard.edu/abs/2018A&A...618A.164S}
}

@ARTICLE{2021MNRAS.501..491V,
	author = {{Van de Sande}, M. and {Walsh}, C. and {Millar}, T.~J.},
	title = "{Chemical modelling of dust-gas chemistry within AGB outflows - III. Photoprocessing of the ice and return to the ISM}",
	journal = {\mnras},
	year = 2021,
	month = feb,
	volume = {501},
	number = {1},
	pages = {491-506},
	doi = {10.1093/mnras/staa3689},
	archivePrefix = {arXiv},
	eprint = {2011.11563},
	primaryClass = {astro-ph.SR},
	adsurl = {https://ui.adsabs.harvard.edu/abs/2021MNRAS.501..491V}
}

@ARTICLE{2022MNRAS.512.1091S,
	author = {{Scicluna}, P. and {Kemper}, F. and {McDonald}, I. and {Srinivasan}, S. and {Trejo}, A. and {Wallstr{\"o}m}, S.~H.~J. and {Wouterloot}, J.~G.~A. and {Cami}, J. and {Greaves}, J. and {He}, Jinhua and {Hoai}, D.~T. and {Kim}, Hyosun and {Jones}, O.~C. and {Shinnaga}, H. and {Clark}, C.~J.~R. and {Dharmawardena}, T. and {Holland}, W. and {Imai}, H. and {van Loon}, J. Th and {Menten}, K.~M. and {Wesson}, R. and {Chawner}, H. and {Feng}, S. and {Goldman}, S. and {Liu}, F.~C. and {MacIsaac}, H. and {Tang}, J. and {Zeegers}, S. and {Amada}, K. and {Antoniou}, V. and {Bemis}, A. and {Boyer}, M.~L. and {Chapman}, S. and {Chen}, X. and {Cho}, S.-H. and {Cui}, L. and {Dell'Agli}, F. and {Friberg}, P. and {Fukaya}, S. and {Gomez}, H. and {Gong}, Y. and {Hadjara}, M. and {Haswell}, C. and {Hirano}, N. and {Hony}, S. and {Izumiura}, H. and {Jeste}, M. and {Jiang}, X. and {Kaminski}, T. and {Keaveney}, N. and {Kim}, J. and {Kraemer}, K.~E. and {Kuan}, Y.-J. and {Lagadec}, E. and {Lee}, C.~F. and {Li}, D. and {Liu}, S.-Y. and {Liu}, T. and {de Looze}, I. and {Lykou}, F. and {Maraston}, C. and {Marshall}, J.~P. and {Matsuura}, M. and {Min}, C. and {Otsuka}, M. and {Oyadomari}, M. and {Parsons}, H. and {Patel}, N.~A. and {Peeters}, E. and {Pham}, T.~A. and {Qiu}, J. and {Randall}, S. and {Rau}, G. and {Redman}, M.~P. and {Richards}, A.~M.~S. and {Serjeant}, S. and {Shi}, C. and {Sloan}, G.~C. and {Smith}, M.~W.~L. and {Suh}, K.-W. and {Toal{\'a}}, J.~A. and {Uttenthaler}, S. and {Ventura}, P. and {Wang}, B. and {Yamamura}, I. and {Yang}, T. and {Yun}, Y. and {Zhang}, F. and {Zhang}, Y. and {Zhao}, G. and {Zhu}, M. and {Zijlstra}, A.~A.},
	title = "{The Nearby Evolved Stars Survey II: Constructing a volume-limited sample and first results from the James Clerk Maxwell Telescope}",
	journal = {\mnras},
	year = 2022,
	month = may,
	volume = {512},
	number = {1},
	pages = {1091-1110},
	doi = {10.1093/mnras/stab2860},
	archivePrefix = {arXiv},
	eprint = {2110.12562},
	primaryClass = {astro-ph.GA},
	adsurl = {https://ui.adsabs.harvard.edu/abs/2022MNRAS.512.1091S}
}

@ARTICLE{2019MNRAS.490.3158C,
       author = {{Cruzal{\`e}bes}, P. and {Petrov}, R.~G. and {Robbe-Dubois}, S. and {Varga}, J. and {Burtscher}, L. and {Allouche}, F. and {Berio}, P. and {Hofmann}, K.-H. and {Hron}, J. and {Jaffe}, W. and {Lagarde}, S. and {Lopez}, B. and {Matter}, A. and {Meilland}, A. and {Meisenheimer}, K. and {Millour}, F. and {Schertl}, D.},
        title = "{A catalogue of stellar diameters and fluxes for mid-infrared interferometry}",
      journal = {\mnras},
         year = 2019,
        month = dec,
       volume = {490},
       number = {3},
        pages = {3158-3176},
          doi = {10.1093/mnras/stz2803},
archivePrefix = {arXiv},
       eprint = {1910.00542},
 primaryClass = {astro-ph.SR},
       adsurl = {https://ui.adsabs.harvard.edu/abs/2019MNRAS.490.3158C}
}

@ARTICLE{2026A&A...711A.159M,
       author = {{Marinho}, L. and {Fonfr{\'\i}a}, J.~P. and {Ag{\'u}ndez}, M. and {Velilla-Prieto}, L. and {Quintana-Lacaci}, G. and {Cernicharo}, J.},
        title = "{The formation radius of HCN in O-rich asymptotic giant branch stars}",
      journal = {\aap},
         year = 2026,
        month = jul,
       volume = {711},
          eid = {A159},
        pages = {A159},
          doi = {10.1051/0004-6361/202660604},
archivePrefix = {arXiv},
       eprint = {2606.19128},
 primaryClass = {astro-ph.SR},
       adsurl = {https://ui.adsabs.harvard.edu/abs/2026A&A...711A.159M}
}

@ARTICLE{2018A&ARv..26....1H,
	author = {{H{\"o}fner}, Susanne and {Olofsson}, Hans},
	title = "{Mass loss of stars on the asymptotic giant branch. Mechanisms, models and measurements}",
	journal = {\aapr},
	year = 2018,
	month = jan,
	volume = {26},
	number = {1},
	eid = {1},
	pages = {1},
	doi = {10.1007/s00159-017-0106-5},
	adsurl = {https://ui.adsabs.harvard.edu/abs/2018A&ARv..26....1H}
}

@ARTICLE{1993AJ....105..595S,
	author = {{Sahai}, Raghvendra and {Bieging}, John H.},
	title = "{Interferometric Observations of Non-Maser SiO Emission From Circumstellar Envelopes of AGB Stars: Acceleration Regions and SIO Depletion}",
	journal = {\aj},
	year = 1993,
	month = feb,
	volume = {105},
	pages = {595},
	doi = {10.1086/116456},
	adsurl = {https://ui.adsabs.harvard.edu/abs/1993AJ....105..595S}
}

@ARTICLE{1985ApJ...292..487J,
	author = {{Jura}, M. and {Morris}, M.},
	title = "{Condensation onto grains in the outflows from mass-losing red giants.}",
	journal = {\apj},
	year = 1985,
	month = may,
	volume = {292},
	pages = {487-493},
	doi = {10.1086/163180},
	adsurl = {https://ui.adsabs.harvard.edu/abs/1985ApJ...292..487J}
}

@ARTICLE{2023AJ....165..229S,
	author = {{Sahai}, Raghvendra and {Stenger}, Benjamin},
	title = "{Faint but Not Forgotten. I. First Results from a Search for Astrospheres around AGB Stars in the Far-ultraviolet}",
	journal = {\aj},
	year = 2023,
	month = jun,
	volume = {165},
	number = {6},
	eid = {229},
	pages = {229},
	doi = {10.3847/1538-3881/acccf2},
	archivePrefix = {arXiv},
	eprint = {2305.07735},
	primaryClass = {astro-ph.SR},
	adsurl = {https://ui.adsabs.harvard.edu/abs/2023AJ....165..229S}
}

@ARTICLE{2026arXiv260816456G,
	author = {{Groenewegen}, M.~A.~T. and {Nanni}, A. and {Boyer}, M.~L. and {Decin}, L. and {Goldman}, S.~R. and {Justtanont}, K. and {Kerschbaum}, F. and {McDonald}, I. and {Olofsson}, H. and {Sahai}, R. and {Sloan}, G.~C. and {van Loon}, J. Th. and {Vlemmings}, W.~H.~T. and {Zijlstra}, A.~A.},
	title = "{CO rotational line emission in very red carbon stars in the Magellanic Clouds}",
	journal = {arXiv e-prints},
	year = 2026,
	month = aug,
	eid = {arXiv:2608.16456},
	pages = {arXiv:2608.16456},
	doi = {10.48550/arXiv.2608.16456},
	archivePrefix = {arXiv},
	eprint = {2608.16456},
	primaryClass = {astro-ph.SR},
	adsurl = {https://ui.adsabs.harvard.edu/abs/2026arXiv260816456G}
}

@ARTICLE{2016A&A...596A..50G,
	author = {{Groenewegen}, M.~A.~T. and {Vlemmings}, W.~H.~T. and {Marigo}, P. and {Sloan}, G.~C. and {Decin}, L. and {Feast}, M.~W. and {Goldman}, S.~R. and {Justtanont}, K. and {Kerschbaum}, F. and {Matsuura}, M. and {McDonald}, I. and {Olofsson}, H. and {Sahai}, R. and {van Loon}, J. Th. and {Wood}, P.~R. and {Zijlstra}, A.~A. and {Bernard-Salas}, J. and {Boyer}, M.~L. and {Guzman-Ramirez}, L. and {Jones}, O.~C. and {Lagadec}, E. and {Meixner}, M. and {Rawlings}, M.~G. and {Srinivasan}, S.},
	title = "{The ALMA detection of CO rotational line emission in AGB stars in the Large Magellanic Cloud}",
	journal = {\aap},
	year = 2016,
	month = nov,
	volume = {596},
	eid = {A50},
	pages = {A50},
	doi = {10.1051/0004-6361/201629590},
	archivePrefix = {arXiv},
	eprint = {1609.09647},
	primaryClass = {astro-ph.SR},
	adsurl = {https://ui.adsabs.harvard.edu/abs/2016A&A...596A..50G}
}

@ARTICLE{2025A&A...704A.341D,
	author = {{Danilovich}, T. and {Samaratunge}, N. and {Mori}, Y.~L. and {Richards}, A.~M.~S. and {Baudry}, A. and {Etoka}, S. and {Montarg{\`e}s}, M. and {Kervella}, P. and {McDonald}, I. and {Gottlieb}, C.~A. and {Wallace}, A. and {Price}, D.~J. and {Decin}, L. and {Bolte}, J. and {Ceulemans}, T. and {De Ceuster}, F. and {de Koter}, A. and {Dionese}, D. and {El Mellah}, I. and {Esseldeurs}, M. and {Gray}, M. and {Herpin}, F. and {Khouri}, T. and {Lagadec}, E. and {Landri}, C. and {Marinho}, L. and {Menten}, K.~M. and {Millar}, T.~J. and {M{\"u}ller}, H.~S.~P. and {Pimpanuwat}, B. and {Plane}, J.~M.~C. and {Sahai}, R. and {Siess}, L. and {Van de Sande}, M. and {Vermeulen}, O. and {Wong}, K.~T. and {Yates}, J. and {Zijlstra}, A.},
	title = "{ATOMIUM: Continuum emission and evidence of dust enhancement from binary motion}",
	journal = {\aap},
	year = 2025,
	month = dec,
	volume = {704},
	eid = {A341},
	pages = {A341},
	doi = {10.1051/0004-6361/202554878},
	archivePrefix = {arXiv},
	eprint = {2504.00517},
	primaryClass = {astro-ph.SR},
	adsurl = {https://ui.adsabs.harvard.edu/abs/2025A&A...704A.341D}
}

@ARTICLE{2026A&A...707A.162K,
	author = {{Khouri}, T. and {Vlemmings}, W.~H.~T. and {Raudales Oseguera}, D.~A. and {Tafoya}, D. and {Olofsson}, H. and {Paladini}, C. and {Maercker}, M. and {Saberi}, M. and {Gorai}, P. and {Danilovich}, T.},
	title = "{A surprisingly large asymmetric ejection from Mira A}",
	journal = {\aap},
	year = 2026,
	month = mar,
	volume = {707},
	eid = {A162},
	pages = {A162},
	doi = {10.1051/0004-6361/202557159},
	archivePrefix = {arXiv},
	eprint = {2602.03159},
	primaryClass = {astro-ph.SR},
	adsurl = {https://ui.adsabs.harvard.edu/abs/2026A&A...707A.162K}
}

@ARTICLE{2025A&A...698A.179D,
	author = {{De Beck}, E. and {Andrews}, H. and {Quintana-Lacaci}, G. and {Vlemmings}, W.~H.~T.},
	title = "{The circumstellar environment around the extreme Galactic red supergiant NML Cygni: Dense, dusty, and asymmetric}",
	journal = {\aap},
	year = 2025,
	month = jun,
	volume = {698},
	eid = {A179},
	pages = {A179},
	doi = {10.1051/0004-6361/202554748},
	archivePrefix = {arXiv},
	eprint = {2505.02643},
	primaryClass = {astro-ph.GA},
	adsurl = {https://ui.adsabs.harvard.edu/abs/2025A&A...698A.179D}
}

@ARTICLE{2008ApJ...673..445F,
	author = {{Fonfr{\'\i}a}, J.~P. and {Cernicharo}, J. and {Richter}, M.~J. and {Lacy}, J.~H.},
	title = "{A Detailed Analysis of the Dust Formation Zone of IRC +10216 Derived from Mid-Infrared Bands of C$_{2}$H$_{2}$ and HCN}",
	journal = {\apj},
	year = 2008,
	month = jan,
	volume = {673},
	number = {1},
	pages = {445-469},
	doi = {10.1086/523882},
	archivePrefix = {arXiv},
	eprint = {0709.4390},
	primaryClass = {astro-ph},
	adsurl = {https://ui.adsabs.harvard.edu/abs/2008ApJ...673..445F}
}

@ARTICLE{2020Sci...369.1497D,
	author = {{Decin}, L. and {Montarg{\`e}s}, M. and {Richards}, A.~M.~S. and {Gottlieb}, C.~A. and {Homan}, W. and {McDonald}, I. and {El Mellah}, I. and {Danilovich}, T. and {Wallstr{\"o}m}, S.~H.~J. and {Zijlstra}, A. and {Baudry}, A. and {Bolte}, J. and {Cannon}, E. and {De Beck}, E. and {De Ceuster}, F. and {de Koter}, A. and {De Ridder}, J. and {Etoka}, S. and {Gobrecht}, D. and {Gray}, M. and {Herpin}, F. and {Jeste}, M. and {Lagadec}, E. and {Kervella}, P. and {Khouri}, T. and {Menten}, K. and {Millar}, T.~J. and {M{\"u}ller}, H.~S.~P. and {Plane}, J.~M.~C. and {Sahai}, R. and {Sana}, H. and {Van de Sande}, M. and {Waters}, L.~B.~F.~M. and {Wong}, K.~T. and {Yates}, J.},
	title = "{(Sub)stellar companions shape the winds of evolved stars}",
	journal = {Science},
	year = 2020,
	month = sep,
	volume = {369},
	number = {6510},
	pages = {1497-1500},
	doi = {10.1126/science.abb1229},
	archivePrefix = {arXiv},
	eprint = {2009.11694},
	primaryClass = {astro-ph.SR},
	adsurl = {https://ui.adsabs.harvard.edu/abs/2020Sci...369.1497D}
}

@ARTICLE{2012Natur.490..232M,
	author = {{Maercker}, M. and {Mohamed}, S. and {Vlemmings}, W.~H.~T. and {Ramstedt}, S. and {Groenewegen}, M.~A.~T. and {Humphreys}, E. and {Kerschbaum}, F. and {Lindqvist}, M. and {Olofsson}, H. and {Paladini}, C. and {Wittkowski}, M. and {de Gregorio-Monsalvo}, I. and {Nyman}, L.-A.},
	title = "{Unexpectedly large mass loss during the thermal pulse cycle of the red giant star R Sculptoris}",
	journal = {\nat},
	year = 2012,
	month = oct,
	volume = {490},
	number = {7419},
	pages = {232-234},
	doi = {10.1038/nature11511},
	archivePrefix = {arXiv},
	eprint = {1210.3030},
	primaryClass = {astro-ph.SR},
	adsurl = {https://ui.adsabs.harvard.edu/abs/2012Natur.490..232M}
}

@ARTICLE{2026arXiv260823695V,
	author = {{Vlemmings}, W. and {Andriantsaralaza}, M. and {Bojnordi Arbab}, B. and {Khouri}, T. and {Lankhaar}, B. and {Planquart}, L. and {De Beck}, E. and {Del Palacio}, S. and {Humphreys}, E. and {Jadlovsky}, D. and {Maercker}, M. and {Olofsson}, H. and {Saberi}, M. and {Siebert}, M. and {Unnikrishnan}, R. and {Wedemeyer}, S. and {Wittkowski}, M.},
	title = "{A giant magnetic eruption from the asymptotic giant branch star Mira A}",
	journal = {arXiv e-prints},
	year = 2026,
	month = aug,
	eid = {arXiv:2608.23695},
	pages = {arXiv:2608.23695},
	doi = {10.48550/arXiv.2608.23695},
	archivePrefix = {arXiv},
	eprint = {2608.23695},
	primaryClass = {astro-ph.SR},
	adsurl = {https://ui.adsabs.harvard.edu/abs/2026arXiv260823695V}
}

@ARTICLE{2015A&A...573L...1O,
	author = {{O'Gorman}, E. and {Vlemmings}, W. and {Richards}, A.~M.~S. and {Baudry}, A. and {De Beck}, E. and {Decin}, L. and {Harper}, G.~M. and {Humphreys}, E.~M. and {Kervella}, P. and {Khouri}, T. and {Muller}, S.},
	title = "{ALMA observations of anisotropic dust mass loss in the inner circumstellar environment of the red supergiant VY Canis Majoris}",
	journal = {\aap},
	year = 2015,
	month = jan,
	volume = {573},
	eid = {L1},
	pages = {L1},
	doi = {10.1051/0004-6361/201425101},
	archivePrefix = {arXiv},
	eprint = {1410.1622},
	primaryClass = {astro-ph.SR},
	adsurl = {https://ui.adsabs.harvard.edu/abs/2015A&A...573L...1O}
}

@ARTICLE{2023A&A...669A..56Q,
	author = {{Quintana-Lacaci}, G. and {Velilla-Prieto}, L. and {Ag{\'u}ndez}, M. and {Fonfr{\'\i}a}, J.~P. and {Cernicharo}, J. and {Decin}, L. and {Castro-Carrizo}, A.},
	title = "{History of two mass loss processes in VY CMa. Fast outflows carving older ejecta}",
	journal = {\aap},
	year = 2023,
	month = jan,
	volume = {669},
	eid = {A56},
	pages = {A56},
	doi = {10.1051/0004-6361/202244396},
	archivePrefix = {arXiv},
	eprint = {2210.16135},
	primaryClass = {astro-ph.SR},
	adsurl = {https://ui.adsabs.harvard.edu/abs/2023A&A...669A..56Q}
}

@article{Li_2026,
	doi = {10.3847/2041-8213/aea794},
	url = {https://doi.org/10.3847/2041-8213/aea794},
	year = {2026},
	month = {sep},
	publisher = {The American Astronomical Society},
	volume = {1009},
	number = {2},
	pages = {L47},
	author = {{Li}, J. and García-Hernández, D. A. and Manchado, A. and Das, Debayan and Cami, J. and Peeters, Els and Sloan, G. C. and Aringer, B. and Bernard-Salas, J. and Bhatt, C. and Clark, Nicholas and Dinerstein, Harriet L. and Gómez-Muñoz, M. A. and Kraemer, Kathleen E. and Matsuura, M. and Sahai, R. and Sterling, N. C. and Volk, Kevin and Wahlgren, G. M. and Zijlstra, A. A.},
	title = {The JWST Proto-PAH Project: Detection of the Methyl Radical CH3 in the Highly Evolved C-rich Object SMP LMC 011},
	journal = {The Astrophysical Journal Letters},
	}

@ARTICLE{2022NatAs...6.1421D,
	author = {{De Marco}, Orsola and {Akashi}, Muhammad and {Akras}, Stavros and {Alcolea}, Javier and {Aleman}, Isabel and {Amram}, Philippe and {Balick}, Bruce and {De Beck}, Elvire and {Blackman}, Eric G. and {Boffin}, Henri M.~J. and {Boumis}, Panos and {Bublitz}, Jesse and {Bucciarelli}, Beatrice and {Bujarrabal}, Valentin and {Cami}, Jan and {Chornay}, Nicholas and {Chu}, You-Hua and {Corradi}, Romano L.~M. and {Frank}, Adam and {Garc{\'\i}a-Hern{\'a}ndez}, D.~A. and {Garc{\'\i}a-Rojas}, Jorge and {Garc{\'\i}a-Segura}, Guillermo and {G{\'o}mez-Llanos}, Veronica and {Gon{\c{c}}alves}, Denise R. and {Guerrero}, Mart{\'\i}n A. and {Jones}, David and {Karakas}, Amanda I. and {Kastner}, Joel H. and {Kwok}, Sun and {Lykou}, Foteini and {Manchado}, Arturo and {Matsuura}, Mikako and {McDonald}, Iain and {Miszalski}, Brent and {Mohamed}, Shazrene S. and {Monreal-Ibero}, Ana and {Monteiro}, Hektor and {Montez}, Rodolfo and {Baez}, Paula Moraga and {Morisset}, Christophe and {Nordhaus}, Jason and {Mendes de Oliveira}, Claudia and {Osborn}, Zara and {Otsuka}, Masaaki and {Parker}, Quentin A. and {Peeters}, Els and {Quint}, Bruno C. and {Quintana-Lacaci}, Guillermo and {Redman}, Matt and {Ruiter}, Ashley J. and {Sabin}, Laurence and {Sahai}, Raghvendra and {Contreras}, Carmen S{\'a}nchez and {Santander-Garc{\'\i}a}, Miguel and {Seitenzahl}, Ivo and {Soker}, Noam and {Speck}, Angela K. and {Stanghellini}, Letizia and {Steffen}, Wolfgang and {Toal{\'a}}, Jes{\'u}s A. and {Ueta}, Toshiya and {Van de Steene}, Griet and {Van Winckel}, Hans and {Ventura}, Paolo and {Villaver}, Eva and {Vlemmings}, Wouter and {Walsh}, Jeremy R. and {Wesson}, Roger and {Zijlstra}, Albert A.},
	title = "{The messy death of a multiple star system and the resulting planetary nebula as observed by JWST}",
	journal = {Nature Astronomy},
	year = 2022,
	month = dec,
	volume = {6},
	pages = {1421-1432},
	doi = {10.1038/s41550-022-01845-2},
	archivePrefix = {arXiv},
	eprint = {2301.02775},
	primaryClass = {astro-ph.SR},
	adsurl = {https://ui.adsabs.harvard.edu/abs/2022NatAs...6.1421D}
}

@ARTICLE{2013ApJ...778L..25C,
	author = {{Cernicharo}, J. and {Daniel}, F. and {Castro-Carrizo}, A. and {Agundez}, M. and {Marcelino}, N. and {Joblin}, C. and {Goicoechea}, J.~R. and {Gu{\'e}lin}, M.},
	title = "{Unveiling the Dust Nucleation Zone of IRC+10216 with ALMA}",
	journal = {\apjl},
	year = 2013,
	month = dec,
	volume = {778},
	number = {2},
	eid = {L25},
	pages = {L25},
	doi = {10.1088/2041-8205/778/2/L25},
	archivePrefix = {arXiv},
	eprint = {1310.7383},
	primaryClass = {astro-ph.GA},
	adsurl = {https://ui.adsabs.harvard.edu/abs/2013ApJ...778L..25C}
}

@ARTICLE{2019A&A...627L...4C,
	author = {{Cernicharo}, J. and {Velilla-Prieto}, L. and {Ag{\'u}ndez}, M. and {Pardo}, J.~R. and {Fonfr{\'\i}a}, J.~P. and {Quintana-Lacaci}, G. and {Cabezas}, C. and {Berm{\'u}dez}, C. and {Gu{\'e}lin}, M.},
	title = "{Discovery of the first Ca-bearing molecule in space: CaNC}",
	journal = {\aap},
	year = 2019,
	month = jul,
	volume = {627},
	eid = {L4},
	pages = {L4},
	doi = {10.1051/0004-6361/201936040},
	archivePrefix = {arXiv},
	eprint = {1906.09352},
	primaryClass = {astro-ph.SR},
	adsurl = {https://ui.adsabs.harvard.edu/abs/2019A&A...627L...4C}
}

@ARTICLE{2025A&A...700L...6P,
	author = {{Pardo}, J.~R. and {Fonfr{\'\i}a}, J.~P. and {Ag{\'u}ndez}, M. and {Tercero}, B. and {Cabezas}, C. and {Velilla-Prieto}, L. and {Gu{\'e}lin}, M. and {de Vicente}, P. and {Cernicharo}, J.},
	title = "{Detection of SiC$_{6}$ in IRC +10216 through Q-band lines}",
	journal = {\aap},
	year = 2025,
	month = aug,
	volume = {700},
	eid = {L6},
	pages = {L6},
	doi = {10.1051/0004-6361/202555902},
	adsurl = {https://ui.adsabs.harvard.edu/abs/2025A&A...700L...6P}
}
\end{document}